\documentclass{vgtc}                          
\pdfoutput=1

\ifpdf
  \pdfoutput=1\relax                   
  \usepackage{graphicx}                
  \DeclareGraphicsExtensions{.pdf,.png,.jpg,.jpeg} 
\else
  \usepackage{graphicx}                
  \DeclareGraphicsExtensions{.eps}     
\fi%

\graphicspath{{figures/}{pictures/}{images/}{./}} 

\usepackage{microtype}                 
\PassOptionsToPackage{warn}{textcomp}  
\usepackage{textcomp}                  
\usepackage{mathptmx}                  
\usepackage{times}                     
\usepackage{cite}                      
\usepackage{tabu}                      
\usepackage{booktabs}                  
\usepackage{amsmath}
\usepackage{graphicx}	
\usepackage{amsmath}	
\usepackage{amssymb}	
\usepackage{booktabs}
\usepackage{times}
\usepackage{microtype}
\usepackage{epsfig}
\usepackage[table,xcdraw,dvipsnames]{xcolor}
\usepackage{caption}
\usepackage{float}
\usepackage{placeins}
\usepackage{color, colortbl}
\usepackage{stfloats}
\usepackage{enumitem}
\usepackage{tabularx}
\usepackage{xstring}
\usepackage{multirow}
\usepackage{xspace}
\usepackage{url}
\usepackage{subcaption}
\usepackage{xcolor}
\usepackage[hang,flushmargin]{footmisc}

\usepackage{graphicx}
\usepackage{amsmath}
\usepackage{amssymb}
\usepackage{booktabs}
\usepackage{array}
\usepackage{multirow}
\usepackage{makecell}
\usepackage{bbding}
\PassOptionsToPackage{normalem}{ulem}
\usepackage{ulem}
\usepackage{arydshln}
\usepackage{xcolor}         
\usepackage[accsupp]{axessibility}
\usepackage{pifont}
\usepackage{combelow}  
\usepackage{hyperref}
\usepackage{graphicx}
\usepackage{amssymb}
\usepackage{xcolor}
\usepackage{pifont}
\usepackage{multirow}

\newcommand{\eg}{e.\,g.\ }

\usepackage{enumitem}
\newlist{questions}{enumerate}{2}
\setlist[questions,1]{label=RQ\arabic*.,ref=RQ\arabic*}
\setlist[questions,2]{label=(\alph*),ref=\thequestionsi(\alph*)}

\onlineid{0}

\vgtccategory{Research}

\title{M2fNet: Multi-modal Forest Monitoring Network on Large-scale \\ Virtual Dataset}

\newcommand{\authors}{
	{\textsuperscript{1}} Yawen \ Lu\ \thanks{e-mail: lu976@purdue.edu}, \ \
	{\textsuperscript{1}} Yunhan \ Huang\ \thanks{e-mail: huan1482@purdue.edu}, \ \
	{\textsuperscript{1}} Su \ Sun\ \thanks{e-mail: sun931@purdue.edu}, \ \
	{\textsuperscript{1}} Tansi \ Zhang\ \thanks{e-mail: zhan4138@purdue.edu}, \ \
	{\textsuperscript{2}} Xuewen \ Zhang\thanks{e-mail: xuewenzhang@meta.com}, \ \
	{\textsuperscript{3}} Songlin \ Fei\thanks{e-mail: sfei@purdue.edu}, \ \
	{\textsuperscript{1}} Victor \ Chen\ \thanks{e-mail: victorchen@purdue.edu} \ \
}

\author{\authors\\
     \scriptsize \textsuperscript{1}Department of Computer Graphics Technology, Purdue University \ \ 
     \scriptsize \textsuperscript{2}Meta, USA  \\
     \scriptsize \textsuperscript{3}Department of Forestry and Natural Resources, Purdue University 
}

\begin{document}
\maketitle

\abstract{Forest monitoring and education are key to forest protection, education and management, which is an effective way to measure the progress of a country's forest and climate commitments. Due to the lack of a large-scale wild forest monitoring benchmark, the common practice is to train the model on a common outdoor benchmark (e.g., KITTI) and evaluate it on real forest datasets (e.g., CanaTree100). However, there is a large domain gap in this setting, which makes the evaluation and deployment difficult. In this paper, we propose a new photorealistic virtual forest dataset and a multimodal transformer-based algorithm for tree detection and instance segmentation. To the best of our knowledge, it is the first time that a multimodal detection and segmentation algorithm is applied to a large-scale forest scenes. We believe that the proposed dataset and method will inspire the simulation, computer vision, education and forestry communities towards a more comprehensive multi-modal understanding.}


\vspace{-1mm}
\CCScatlist{
  \CCScatTwelve{Computing methodologies}{Modeling and simulation}{Simulation support systems}{Simulation environments};
  \CCScatTwelve{Computing methodologies}{Artificial intelligence}{Computer vision}{Image segmentation / Object detection.}
}





\section{Introduction} 
Forest monitoring and simulation (e.g., tree detection, counting, segmentation and reconstruction) are  highly active field in forestry, education and computer science ~\cite{itakura2020automatic, luo2023early, firoze2023tree, grondin2023tree}. Existing datasets for natural forest monitoring (e.g., FinnWoodlands~\cite{lagos2023finnwoodlands}, FOR-instance~\cite{puliti2023instance}) are limited in scale and diversity due to the high costs and logistical challenges inherent in collecting ample annotated data within natural forest environments. As manual annotation relies on human labelers, it also inevitably introduces subjective errors, biases, and noises into the ground-truth labels, especially for trees with limited or complicated features. Given these limitations, an ideal dataset would circumvent the need for manual annotation. Indeed, it is possible to automatically generate photo-realistic training data with error-free labels by utilizing large-scale procedural synthesis.

In this project, we propose a multi-modal forest monitoring network (\textit{M2fNet}) and train the model on the newly simulated virtual forest dataset to evaluate its performance (Fig.~\ref{fig:intro_fig}). To rigorously validate the utility of the proposed dataset and model architecture, we benchmarked instance-level detection and segmentation methods with and without the inclusion of the simulated data, and compared our approach against recent baseline methods. 
Our project page and full dataset will be available at \url{https://forestvrw.github.io/M2fNet/}.

Leveraging our proposed data synthesis and learning framework, interdisciplinary researchers from diverse backgrounds can utilize this system for collaborative analysis of multiple tree attributes (e.g., density, species, etc.). In addition to quantitative benchmarking on segmentation IoU metrics, we conducted a user study engaging domain experts across computer graphics, computer science, design, and forestry. Results validate the efficacy of the overall pipeline for supporting flexible forest monitoring and interactions. The remainder of this paper is organized as follows:
\begin{itemize}
\item Section 2 provides relevant research and core methodological background.
\item Section 3 presents our virtual forest dataset generation workflow, the multi-modal forest monitoring network (\textit{M2fNet}) model architecture, and evaluation protocols.
\item Section 4 highlights potential applications enabled by our approach.
\item Section 5 discusses the future work.
\item Section 6 concludes with a summary of key contributions.
\end{itemize}

\section{Background} 
This literature survey provides an overview of important advances in forest measurement and algorithm training. More specifically, we first summarize the work related to forest measurement and surveying, and then focus specifically on the introduction of vision-based tree detection and segmentation algorithms. 

\subsection{Forest measurement and surveying}

Precision forestry relies on accurate measurements of forest biometrics (e.g., diameter, height, form, etc.). Conventional field-based forest inventories are limited in spatial coverage and can be prohibitively time- and labor-intensive~\cite{chave2006measuring}. Recent studies have sought to automate the acquisition and quantification of key forest biometrics through remote sensing techniques.

Airborne laser scanning (ALS) has been extensively used for area-based estimation of aboveground biomass (AGB) using regression models on metrics like canopy height~\cite{vehmas2011airborne, yin2022three}. While verified against field measurements, ALS coverage can remain spatially sparse. Satellite imaging provides complementary wide-area spectral data, which can predict AGB via empirical model fitting~\cite{hansen2013high, avitabile2016integrated}. However, resolution limits the diagnosis of fine-grained forest properties. To capture intra-stand details, terrestrial laser scanning (TLS), structure-from-motion (SfM), and neural radiance fields (NeRF) are alternative approaches to construct 3D point clouds of sample plots~\cite{liang2016terrestrial, nasiri2021unmanned, lu20213d, ling2023dl3dv}. Though robust, plot-based sampling suffers from generalization and edge artifacts~\cite{nasi2015using}. Meanwhile, unmanned aerial vehicles (UAVs) enable rapid, flexible surveys, but remain constrained by flight endurance and line-of-sight restrictions~\cite{puliti2015inventory, torresan2020individual}.

Our work contributes a large-scale virtual forest data by generating dense, perfectly annotated forest data, which can be used for training on various deep learning models. We demonstrate applications from individual tree delineation to property prediction using diverse synthesized sensing modalities.

\subsection{Vision-based Tree Detection and Segmentation}

Reliable detection and segmentation of individual trees from imagery is a fundamental task in forest monitoring. Previous work in tree crown segmentation relies on hand-crafted features and morphological processing on color aerial imagery~\cite{culvenor2002tida, pouliot2002automated}. With the advent of deep convolutional neural networks (CNNs), data-driven approaches have demonstrated superior performance.

For tree detection, recent methods fine-tune object detection networks such as Faster R-CNN using RGB imagery to generate 2D bounding boxes around tree crowns~\cite{santos2019assessment, luo2022individual}. Light detection and ranging (LiDAR) point clouds have also been used to detect individual trees represented as 3D bounding cylinders~\cite{dai2018new}. Meanwhile, semantic segmentation formulations can produce pixel-wise masks that label tree trunks and canopy extensions in RGB-D data~\cite{kharroubi2022three,lu2023label}.

However, existing datasets for training and evaluating such models are limited in diversity, scalability, and annotation quality~\cite{kaartinen2012international, wu2016individual}. Here, we introduce a large-scale photorealistic synthetic forest rendering engine to automatically generate dense ground-truth data for tree detection and segmentation tasks. We demonstrate the value of our simulation-to-reality paradigm on benchmark models.

\section{Design of the Simulated Dataset} 
\label{sec:dataset}

The manual annotation of vegetation in natural environments is often quite time-intensive and painful, particularly in densely forested areas.
Furthermore, delineating tree boundaries and accounting for occlusion inevitably introduces some errors. To deal with these challenges, we have constructed a simulated virtual forest dataset to efficiently train machine learning algorithms for forestry tasks in a seamless manner. The dataset incorporates high-fidelity, photorealistic models of 19 tree species (\eg Bucida Buceras, Conocarpus Erectus, Eucaliptus Gunni, Melia Azedara, Populus Alba, Quercus Robur, Shinus Molle, Tilia, etc.) rendered in the Unreal Engine, with customizable terrain (existing landscape models or from Unreal Engine landscape editor), weather (dynamic weather like sunset, rain, etc) and lighting (\eg directional, sky, spot, point and rect light) parameters. 

\subsection{Dataset Creation}

\begin{figure}[htb]
\centering
\includegraphics[width=0.99\linewidth, height=5.0cm]{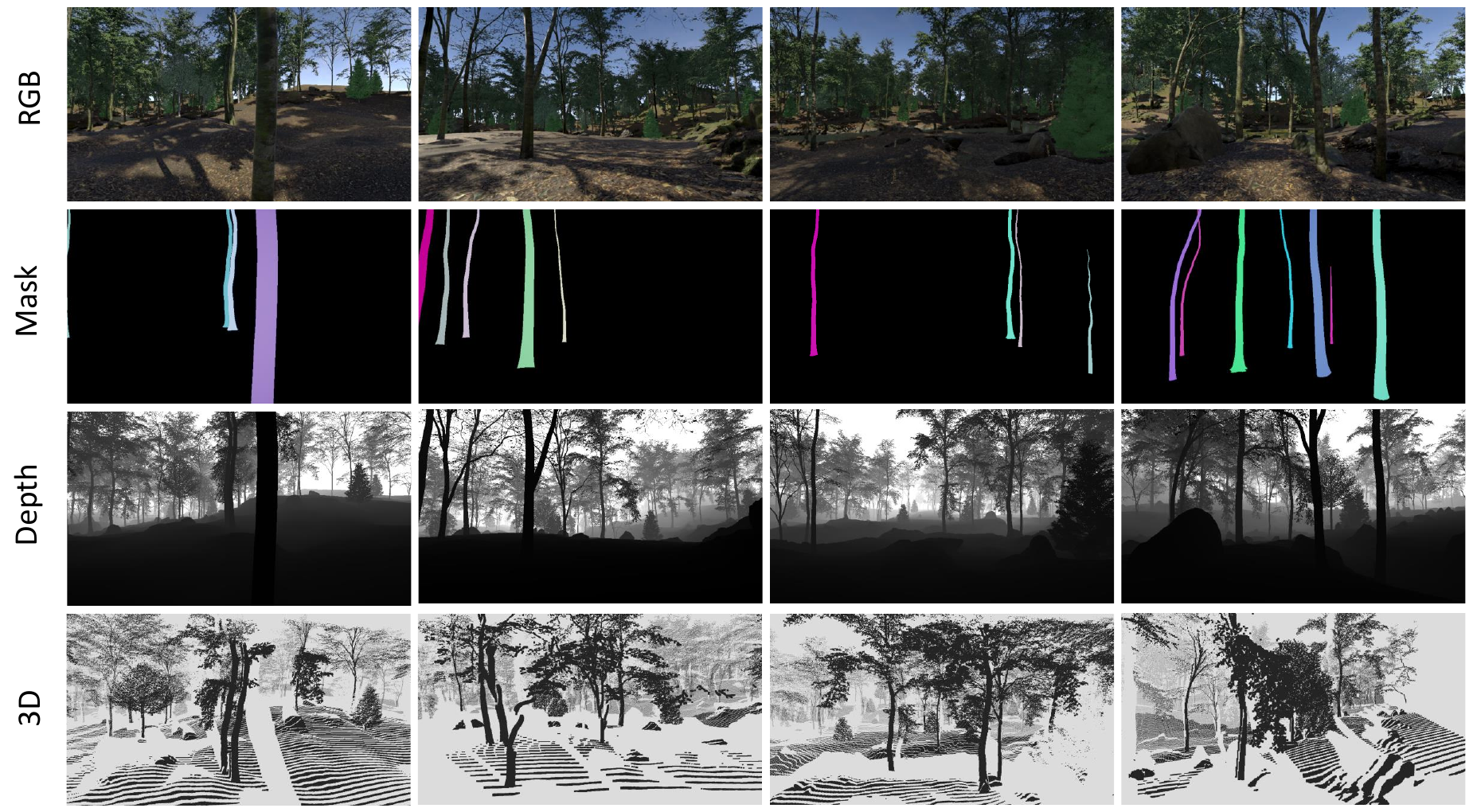}
\vspace{-1mm}
\caption{Examples of our simulated data. Top to bottom: Rendered RGB image, masked image, scene depth, and LiDAR point cloud.
\label{fig:sample}}
\vspace{-2mm}
\end{figure}

In order to precisely segment the regions of interest of the tree trunks from the high-quality tree models collected, Autodesk Maya was utilized to separate each tree model in FBX or OBJ format into its constituent trunk, branch, and leaf components. Only the trunk portions were rendered for the segmentation camera. Individual tree models of different species were populated within the designated terrain map with randomly assigned trunk densities and locations. Within this same terrain, a set of camera trajectories were defined, and the multi-sensor camera was moved along these trajectories to obtain synchronized multimodal collections (e.g., RGB, depth, mask) at each timestamp. This ensured proper alignment of the multimodal image sets. As a post-processing step to generate the COCO annotation files, OpenCV was employed to automatically detect object contours and bounding boxes, which were then exported as the final outputs. As a result, 50,000 video frames containing 3,000 for testing were obtained. Figure \ref{fig:sample} gives some example data from the generated dataset.

\begin{figure}[htb]
\centering
\includegraphics[width=0.99\linewidth, height=5.0cm]{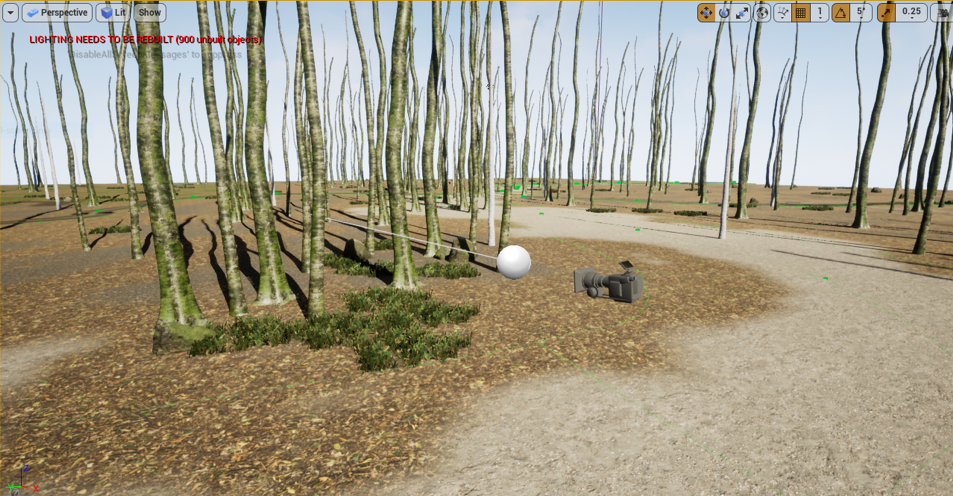}
\vspace{-1mm}
\caption{Project setup for automatic data generation. RGB, depth and semantic cameras are used simultaneously to render forest scenes.
\label{fig:overview}}
\vspace{-3mm}
\end{figure}

The Figure \ref{fig:overview} showcases our project setup based on Unreal Engine 4 for automated data generation within a forest environment. The virtual landscape features adjustable elements like trees, stones, and grass. A camera is in place to capture the scene, which is key for creating diverse datasets. The interface indicates that lighting needs to be rebuilt as pending update to enhance realism. This setup facilitates the generation of varied scenarios for simulation purposes.

\begin{figure}[htb]
\centering
\includegraphics[width=0.99\linewidth, height=4.8cm]{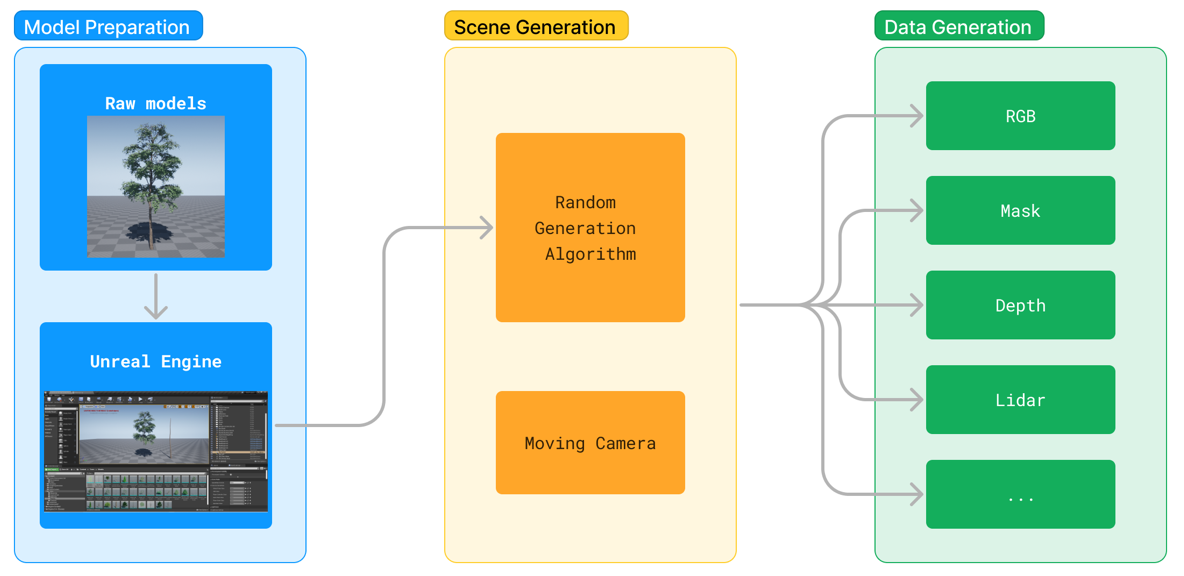}
\vspace{-1mm}
\caption{Three-stage data generation pipeline using Unreal Engine, including tree model preparation, scene generation and data rendering.
\label{fig:pipeline}}
\vspace{-0.1in}
\end{figure}

The Figure \ref{fig:pipeline} outlines a three-step data generation process using Unreal Engine. In the Model Preparation phase, raw models are imported into Unreal Engine for setup. Scene Generation follows, utilizing a random generation algorithm and a moving camera within the environment to create diverse scenarios. Finally, the Data Generation step outputs various data types, such as RGB images, masks, depth maps, and LiDAR. This structured approach enables the creation of a comprehensive dataset for simulation and analysis.

\begin{figure}[htb]
\centering
\includegraphics[width=0.93\linewidth, height=7.3cm]{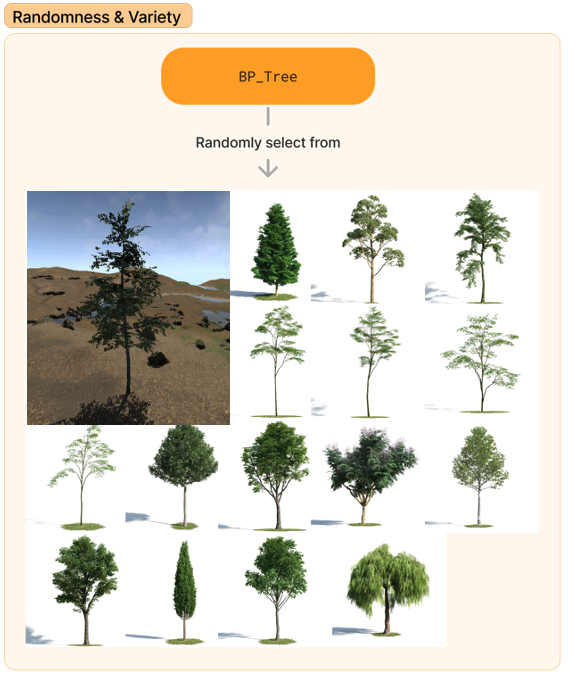}
\vspace{-1mm}
\caption{We design a BP\_Tree class for randomly selecting tree species and models from our collected library.
\label{fig:trees}}
\vspace{-3mm}
\end{figure}

The Figure \ref{fig:trees} displays a selection process for tree models in a simulation environment, showcasing the variety available for scene generation. The blueprint, BP\_Tree, is designed to randomly select from an array of tree species, ensuring randomness and diversity in the virtual landscape. This collection of different tree models allows for a more realistic and varied representation of a forest ecosystem within the simulation.

\subsection{Overview of M2fNet} 

An overview of the proposed multi-modal forest monitoring network is shown in Fig. \ref{fig:intro_fig}. The proposed \textit{M2fNet} uses two independent encoders, which are hierarchical vision transformers with shifted windows (Swin)~\cite{liu2021swin}, to extract RGB and depth features. Considering that RGB and depth images have different emphasis on feature representation, both of which will be helpful for instance-level segmentation and detection. As the encoding stage increases, the features that can be extracted become more advanced. We then pass the features from different domains using a fusion module, which consists of a concatenation, a layer normalization, and a convolution layer. The output of the RGB encoder is added to the output of the fusion module to improve the segmentation performance on small trees and fine tree boundaries.

The fused features are then fed into a pixel decoder to transform the low-resolution features into high-resolution feature embeddings. Similar to Mask2former~\cite{cheng2022masked}, we employ a query-based Transformer decoder with masked attention blocks to learn the semantic-level features from the interested tree objects. Specially, the tree query $Q_{query} \in \mathbb{R}^{N \times d}$ is updated by collecting the local tree features and context information within the foreground regions through self-attention. To deal with small and distant tree instances, high-resolution features from the pixel decoder are fed into the Transformer decoder layer. In this way, after obtaining the aggregated output query $Q_{out}$ and the high-resolution feature $Z$ from the pixel decoder, we predict the tree masks $\mu$ by multiplying the outputs from the two residual structures and adding simple linear layers for the different prediction heads. The predicted mask can be formulated as: $\mathcal{M} = MLP(Z \cdot Q_{out})$, where $MLP$ is the multi-layer projection.

The loss is a combination of the binary cross-entropy loss and the dice loss for mask prediction and the smooth L1 loss for box regression. The total loss can be provided as:

\begin{equation}
L_{final} = \lambda_{bce}L_{bce} + \lambda_{dice}L_{dice} + \lambda_{box}L_{box}
\label{eq:loss}
\end{equation}

\subsection{Metrics and Evaluations}

\begin{figure}[htb]
\centering
\includegraphics[width=0.92\linewidth, height=4.2cm]{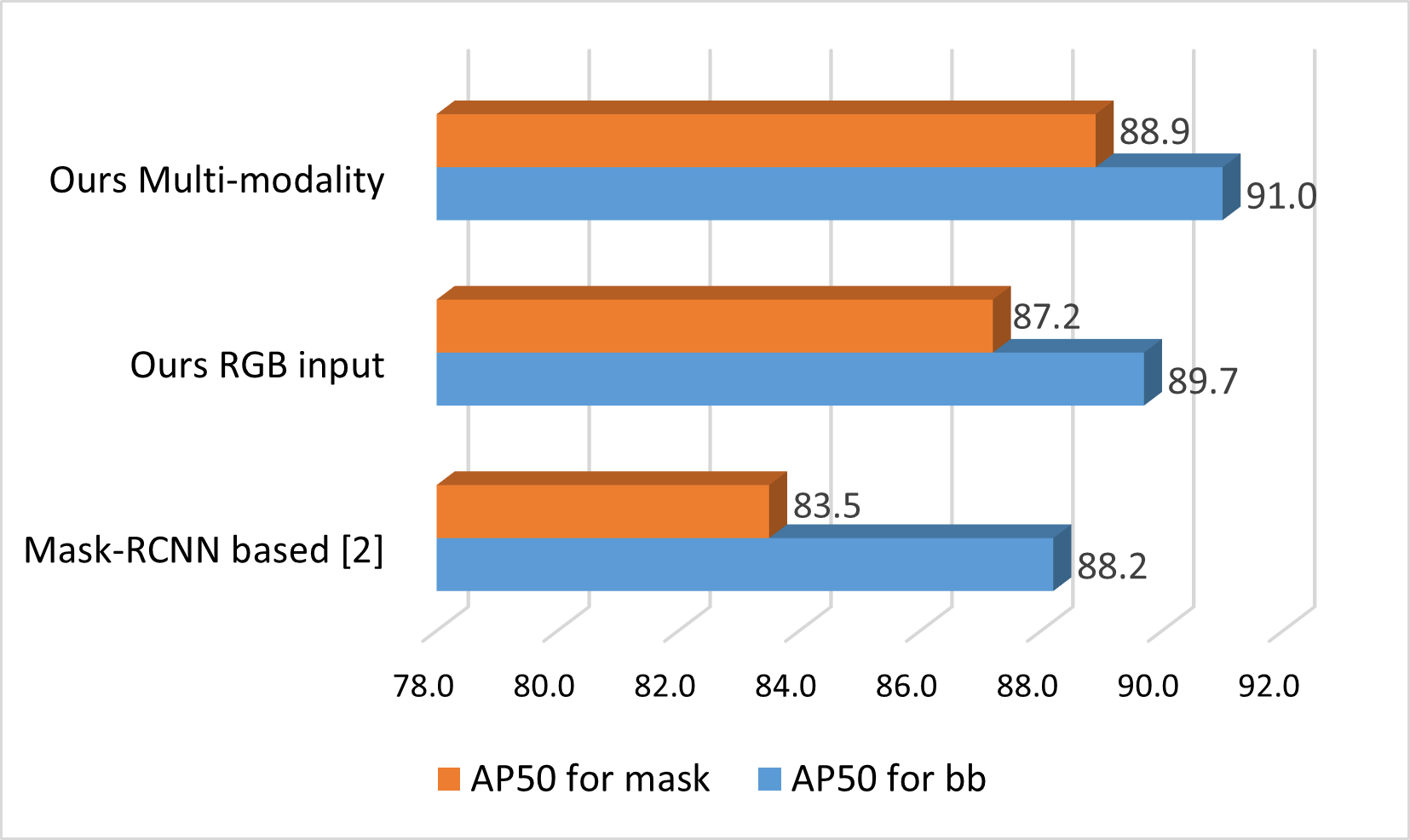}
\vspace{-1.5mm}
\caption{Segmentation and detection evaluation on 100 real images collected in Canadian forests~\cite{grondin2023tree}. The algorithms are greatly improved by incorporating the dataset we created in pre-training.
\label{fig:evaluation}}
\vspace{-4mm}
\end{figure}

To quantify the efficacy of the proposed \textit{M2fNet} architecture using the created simulation data, we employ standard evaluation metrics including mean average precision (mAP) at an IoU threshold of 50\%. As evident in the preliminary results depicted in Figure~\ref{fig:evaluation}, employing our dataset for training precipitates considerable improvements in real-world segmentation and detection performance, especially for tree trunk masks. Furthermore, the multi-modal \textit{M2fNet} outperforms networks relying solely on RGB imagery as input. Collectively, these outcomes underscore the potential for advancing machine learning algorithms, including low-level vision and multimodal frameworks, in unstructured forest environments and other analogous scenarios, by capitalizing on synthetic, controllable datasets.

In addition to the quantitative evaluation described in the previous section, we also conducted a user study aimed at investigating whether the simulated forestry scenes are realistic enough to support as a strong and useful pre-training for subsequent forestry education and research. While user study for measuring simulation system has been approached in VR training systems in the past~\cite{guo2023understanding,jin2023development}, to the best of our knowledge so far has rarely been used in the evaluation of forestry simulator. More specifically, a different level of simulated forest scenes ranging from the most fake to the most realistic (the real-world captured forestry) has been included, covering a broad
range of situations to be investigated in our user study. 15 participants with different backgrounds (computer graphics, design, engineering, forestry) are invited in the study to evaluate the two questions 1 to 5, where the questionnaire for measuring the fidelity of our simulated data can be found in Table~\ref{tab:user}.

\begin{table}[htb]
\centering
\begin{tabular}{ll}
\textbf{RQ1}: Please rate your sense of the fidelity of the rendered images \\
from the simulated forest environment. Which type did you think\\
the reference image came closest to?
\\
\textit{1 - Most fake; 5 - Most realistic} \\
\\
\textbf{RQ2}: How do you feel when the virtual forest scenes become real \\
for you? \\  
\textit{1 - Not at all; 5 - Very much}
\end{tabular}
\vspace{-2mm}
\caption{List of research questions to measure the fidelity of the simulated data.}
\label{tab:user}
\vspace{-4mm}
\end{table}

\section{Usage Scenarios}
\label{sec:usage}

In addition to tree detection and segmentation, the proposed \textit{M2fNet} framework and simulated dataset facilitate the training and evaluation of data-intensive algorithms for various forestry analytical tasks. By leveraging large-scale simulated data, models can mitigate overfitting issues commonly encountered when training on limited real data. Furthermore, the physically-based ground truth generation process provides highly credible labels, enabling robust quantitative analysis and validation of model performance across multiple forestry understanding and education objectives.

\noindent \textbf{Forest DBH Measurement}: 
Diameter at breast height (DBH) measurement constitutes a fundamental metric within forestry practices. Utilizing RGB and/or depth imagery, \textit{M2fNet} employs a multi-modal Transformer architecture trained on the proposed simulated dataset for instance segmentation of individual trees. DBH estimation is subsequently performed by leveraging the predicted segmentation masks in conjunction with depth maps. In contrast to conventional manual techniques (e.g., tape measurements) that often introduce noise into the DBH labels, our framework derives accurate ground truth DBH from 3D tree reconstructions. This facilitates a robust quantitative analysis and validation of the model's DBH predictions, thereby enhancing the overall reliability and validity of the results.

\noindent \textbf{Video-level Tree Tracking}: 
Advances in machine learning allow automated video analysis to identify and track individual trees in wild forest environments. Segmentation results from \textit{M2fNet} can isolate each tree crown from the images in video frames. Tracking algorithms can then match the segmented tree regions across successive frames, tracking a single tree over time even as the camera angle changes. This allows long-term monitoring of specific trees. Researchers can use tracking to build complex tree inventories, measure growth metrics such as crown diameter and height, and detect subtle changes in foliage that may indicate disease. The tree-level data can be scaled up to assess forest population dynamics and health trends. Video-level tree monitoring overcomes many of the limitations of manual surveys, such as small sample size limitations. It also reduces costs and labor compared to field measurements.

\noindent \textbf{Forest Mapping and Reconstruction}: Mapping and reconstruction of forests is imperative for obtaining detailed data on tree species distribution, spatial coordinates, and health indicators. However, prevalent forestry inventory techniques, particularly in large-scale forests and those with dense canopy cover obstructing global navigation satellite systems, often cannot acquire high-quality, reliable measurements. Our simulated dataset offers a robust solution by providing meticulously-recorded camera parameters across trajectory viewpoints and dense point clouds aggregated from forest scenes. This dataset promotes the development and assessment of computational algorithms. Furthermore, the intentionally controlled and varied forest simulation settings (e.g., landscapes and lighting) present in the dataset enable more sophisticated algorithmic comprehension and interpretation of intricate real-world forest environments.

\vspace{-1mm}
\section{Future work}
In the emerging field of digital forestry, our work develops a pioneering multi-modal forest monitoring network leveraging our large-scale photo-realistic simulations. We constructed a large-scale virtual forest dataset using the Unreal Engine, enabling the generation of synthetic data across modalities like LiDAR, spectral imagery, and semantics. This allows training data-hungry deep learning models for tasks like tree detection, species classification, biomass estimation, and forest mapping. Our simulated forest system provides an efficient and scalable platform to prototype and validate AI systems for precision forestry. Looking ahead, such digital twins of forests could transform forestry education, allowing immersive training through simulations. Our virtual forest system is a stepping stone toward digital replicas of natural forests, unlocking new capabilities in forest monitoring and sustainable forest management. By bridging the physical and virtual worlds, this research direction holds promise for understanding complex forest dynamics.

\vspace{-1mm}
\section{Limitations and social impacts}
\label{sec:social}
The methods developed on our virtual forest dataset could generalize well to other natural environments such as mountains and beaches. However, specific outdoor environments such as streets and indoor environments share a large gap in visual appearance with our training focus, and therefore does not perform well. A possible solution will be to adapt to different simulated environments such as cities and rooms. The proposed method assumes RGB or RGB with depth modalities as input, however, it may not be limited to visual modalities only, but can be extended to others, such as text. 

For social impact, the virtual datasets may not accurately represent the diversity of real forests in terms of species composition and ecosystem variability, which may lead to biased decision-making in real-world applications. In addition, the use of virtual datasets may inadvertently neglect social aspects of forest management, such as impacts on local communities and indigenous knowledge.

\vspace{-1mm}
\section{Conclusion}
\label{sec:conclusion}
In this work, we propose a new simulated large-scale forest dataset and a multimodal forest monitoring approach. The dataset contains high-quality and photorealistic rendering data with diverse and controllable properties, taking advantage of the powerful Unreal engine. The developed multimodal forest monitoring network allows precise tree detection and segmentation, enabling a wide range of forestry applications, including tree counting, localization, measurement and 3D mapping. According to the evaluation, using our new dataset and pipeline, the detection and segmentation performance lead to a considerable increase in accuracy.
We hope that our efforts could facilitate further research and progress in the fields of computer vision, forestry, computer graphics, and educational training.

\vspace{-1mm}
\section{Acknowledgments}
This ongoing work is supported by the U.S. Department of Agriculture (USDA) under grant No. 20236801238992.

\bibliographystyle{abbrv-doi}

\bibliography{template}
\end{document}